# Beam mask and sliding window-facilitated deep learning-based accurate and efficient dose prediction for pencil beam scanning proton therapy

Running title: Deep learning-based dose prediction

Lian Zhang, PhD[1*], Jason M. Holmes, PhD[1*], Zhengliang Liu, MS[2], Sujay A. Vora, MD[1], Terence T. Sio, MD, MS[1], Carlos E. Vargas, MD[1], Nathan Y. Yu, MD[1], Sameer R. Keole, MD[1], Steven E. Schild, MD[1], Martin Bues, PhD[1], Sheng Li, PhD[3], Tianming Liu, PhD[2], Jiajian Shen, PhD[1], William W. Wong, MD[1], Wei Liu, PhD[1]

[1]Department of Radiation Oncology, Mayo Clinic, Phoenix, AZ 85054, USA

[2]Department of Computer Science, University of Georgia, Athens, GA 30602, USA

[3]Department of Data Science, University of Virginia, Charlottesville, VA 22903, USA

[*]Co-first authors who contribute to this paper equally

Corresponding author: Wei Liu, PhD, Professor of Radiation Oncology, Department of Radiation Oncology, Mayo Clinic Arizona, 5777 E. Mayo Boulevard, Phoenix, AZ 85054; e-mail: Liu.Wei@mayo.edu.

**Acknowledgments**

This research was supported by the National Cancer Institute (NCI) Career Developmental Award K25CA168984, Arizona Biomedical Research Commission Investigator Award, the Lawrence W. and Marilyn W. Matteson Fund for Cancer Research, and the Kemper Marley Foundation.

**Conflicts of Interest Notification**

Terence T. Sio provides strategic and scientific recommendations as a member of the Advisory Board and speaker for Novocure, Inc., Catalyst Pharmaceuticals, Inc. and Galera Pharmaceutics, which are not in any way associated with the content presented in this manuscript.

**Ethical considerations**

This research was approved by the Mayo Clinic Arizona institutional review board (IRB, 13-005709). The informed consent was waived by IRB protocol. Only CT image and dose-volume data were used in this study. All patient-related health information was removed prior to the analysis and publication of the study.

**Abstract:**

Background: Accurate and efficient dose calculation is essential for on-line adaptive planning in proton therapy. Deep learning (DL) has shown promising dose prediction results in photon therapy. However, there is a scarcity of DL-based dose prediction methods specifically designed for proton therapy. Successful dose prediction method for proton therapy should account for more challenging dose prediction problems in pencil beam scanning proton therapy (PBSPT) due to its sensitivity to heterogeneities.

Purpose: To develop a DL-based PBSPT dose prediction workflow with high accuracy and balanced complexity to support on-line adaptive proton therapy clinical decision and subsequent replanning.

Methods: PBSPT plans of 103 prostate cancer patients (93 for training and the other 10 for independent testing) and 83 lung cancer patients (73 for training and the other 10 for independent testing) previously treated at our institution were included in the study, each with CTs, structure sets, and plan doses calculated by the in-house developed Monte-Carlo dose engine (considered as the ground truth in the model training and testing). For the ablation study, we designed three experiments corresponding to the following three methods: 1) Experiment 1, the conventional region of interest (ROI) (composed of targets and organs-at-risk (OARs)) method. 2) Experiment 2, the beam mask (generated by raytracing of proton beams) method to improve proton dose prediction. 3) Experiment 3, the sliding window method for the model to focus on local details to further improve proton dose prediction. A fully connected 3D-Unet was adopted as the backbone. Dose volume histogram (DVH) indices, 3D Gamma passing rates with a criterion of 3%/3mm/10%, and dice coefficients for the structures enclosed by the iso-dose lines between the predicted and the ground truth doses were used as the evaluation metrics. The calculation time for each proton dose prediction was recorded to evaluate the method's efficiency.

Results: Compared to the conventional ROI method, the beam mask method improved the agreement of DVH indices for both targets and OARs and the sliding window method further improved the agreement of the DVH indices (for lung cancer, CTV D98 absolute deviation: 0.74±0.18 vs. 0.57±0.21 vs. 0.54±0.15 Gy[RBE], ROI vs. beam mask vs. sliding window methods, respectively). For the 3D Gamma passing rates in the target, OARs, and BODY (outside target and OARs), the beam mask method can improve the passing rates in these regions and the sliding window method further improved them (for prostate cancer, targets: 96.93%±0.53% vs. 98.88%±0.49% vs. 99.97%±0.07%, BODY: 86.88%±0.74% vs. 93.21%±0.56% vs. 95.17%±0.59%). A similar trend was also observed for the dice coefficients. In fact, this trend was especially remarkable for relatively low prescription isodose lines (for lung cancer, 10% isodose line dice: 0.871±0.027 vs. 0.911±0.023 vs. 0.927±0.017). The dose predictions for all the testing cases were completed within 0.25s.

Conclusions: An accurate and efficient deep learning-augmented proton dose prediction framework has been developed for PBSPT, which can predict accurate dose distributions not only inside but also outside ROI efficiently. The framework can potentially further reduce the initial planning and adaptive replanning workload in PBSPT.

# 1. Introduction

Pencil beam scanning proton therapy (PBSPT) is a highly adaptable modern beam delivery technique that reduces the dose to healthy tissues as compared to x-ray radiation modalities while not requiring compensators or apertures in most cases[1-13]. PSBPT, however, is more sensitive to range and setup uncertainties in clinical practice as compared to x-ray modalities[14-22]. During the fractionated course of treatment using PBSPT, the patient's anatomical structure tends to change due to weight loss or tumor shrinkage and there may also be random or systemic range and positioning uncertainties[23-31]. As a result, the actual dose received by the patient may deviate from the initially planned dose. To compensate for the potential range and positioning uncertainties, robust optimization is adopted in clinical practice[32-51], however, robust optimization does not account for anatomical changes that may happen during the course of treatment[24].

Adaptive radiotherapy (ART) is a technique designed to address inter-fractional anatomical changes by periodically re-imaging the patient, contouring the new images (or propagating the prior ones), determining whether the original treatment plan is still adequate, and subsequently redoing the treatment planning if not[25, 26, 52]. Conventional ART techniques are costly, requiring a high clinical workload as well as potentially placing a greater burden on the staffs[53-55]. Additionally, because the patient anatomy may change on daily or hourly time scales, for example due to bladder or gas fill, conventional ART techniques may be inadequate[56, 57]. Ideally, a patient would be imaged, their imaging contoured, and the treatment planning performed (if necessary) within minutes, thereby mitigating anatomical changes on these timescales, known as "online" ART[25, 26, 58]. With recent advances in AI-based auto-contouring, able be performed adequately in minutes, the major bottleneck in achieving online ART today, is long dose calculation times, especially in the context of Monte Carlo-based robust optimization[25, 59, 60].

Recently, deep learning (DL) networks have been introduced for dose prediction for photon-based modalities[61]. AI models have the distinct advantage over conventional dose calculation techniques in that they are extremely fast, typically less than 1 second[62, 63]. Fan et al. proposed a Residual Network to predict the dose distribution for intensity-modulated radiation therapy (IMRT) in head-and-neck cancer patients and achieved a clinically acceptable dose distribution for most test cases[64]. Nguyen et al. used a U-net architecture for dose prediction in prostate cancer patients treated with photon therapy and achieved a comparable dose distribution with the dose calculated by conventional means[65]. Other studies have reported similar results using DL for tomotherapy and stereotactic body radiation therapy[66, 67].

Although there have been numerous studies on the use of DL for dose prediction in photon therapy, further research is still needed to predict optimal PBSPT dose distributions. Predicting dose for PBSPT is challenging due to the intrinsic sensitivity of proton dose distribution to anatomical heterogeneities[14]. Dose distributions for PBSPT are therefore highly varied from one patient to another, even for similar treatment sites, as compared with photon-based treatments.

DL-based dose prediction methods specifically for proton therapy are still scare in the literatures compared to photon therapy[68-75]. In this study we aimed to develop a DL-based approach for PBSPT dose prediction to support on-line adaptive proton therapy decision making and potentially the following-up adaptive re-planning. We developed a 3D U-net model using a beam mask to improve the model performance instead of the spot map related information. The beam mask can be generated in a straightforward manner, depending only on the beam angles, making it more practical and deployable in clinical scenario. To further improve the dose prediction accuracy, especially for the low dose regions outside the targets and organs-at-risk (OARs) (defined as regions of interest (ROI)), we proposed a novel 'sliding window' method to have the network concentrate on the local details. Compared to previously published reports, our proposed approach has higher PBSPT dose prediction accuracy and simpler implementation, which is more practical for routine clinical use.

## 2. Material and Methods

2.1 Patient data

For training and testing data, we retrospectively selected 103 prostate cancer and 83 lung cancer patients treated with PBSPT at our institution. For each disease site, 10 cases were randomly selected as the testing group, 5 cases were randomly selected as the validation group during model training, and the others were used for model training. the contours for the target and OARs were examined and approved by experienced radiation oncologists. Prostate cases were prescribed with a dose of 70.0 Gy[RBE] in 28 fractions while lung cases were prescribed with a dose ranging from 50.0 Gy[RBE] to 60.0 Gy[RBE] with 25 to 30 fractions respectively. Treatment plans were designed for each patient based on his/her anatomy using different beam angles and configurations. The plan doses for all the patients included in this study were generated using the our in-house treatment planning system (TPS), Shiva[20, 25, 47, 48, 76-81], which utilizes a Virtual Particle Monte Carlo (VPMC) dose engine[80]. Dose volume constraints for both prostate and lung cases were determined based on our institution clinical standards.

2.2 Data preprocessing

First, the CTs, structure set, and VPMC-calculated dose DICOM files for each patient were extracted and converted to 3D matrices. All the 3D matrices were first resampled to a 2.5 mm grid from the original resolution, and then rigid-registered to a reference case selected from the corresponding site to align the data. The CT HU number and the dose matrices were normalized to have a mean value of 0 and a variance of 1, respectively. We used a box with a dimension of $350 \times 450 \times 550$ centered on the target to crop to ensure all the regions potentially influencing the dose distribution would be included in the model training for all the training and testing data. Zero padding was adopted if the dimension of the processed matrices is less than the cropping box.

A volumetric binary bitmask was then created for each ROI, setting the value to 1 for voxels inside the contour and 0 for those outside the contour. To distinguish the ROI, each ROI will be assigned a specific integer number. For prostate, ROI include CTV, bladder, spaceOAR, left/right femoral head, penile bulb, and rectum. For lung, ROI include CTV, spinal cord, spinal cord prv, esophagus, heart, and total lung. In summary, after preprocessing, the aligned and cropped data for each patient was represented by the CT matrices, contour mask matrices, and dose matrices.

2.3 Beam mask generation

Since the beam paths significantly impacted dose distributions in PBSPT, we enhanced the beam mask method for dose prediction in photon therapy proposed by Peng et al. that provided the DL model with additional beam path information[82]. In addition to the CT and structure bitmask inputs, our 3D beam mask generation method involved extracting beam path information from the radiotherapy treatment plan. We produced the 3D beam masks by assigning a value of 1 to voxels within the beam paths and 0 outside of them by raytracing of the proton beams.

Several enhancements to the beam mask were proposed to improve the dose prediction accuracy for proton therapy as follows. To enhance the dose prediction accuracy at the target distal edge region, we expanded the beam mask margin by 3 mm along the beam direction. For the multi-beam overlapped regions outside the target, the assigned beam mask value will be set to be the value of the sum of the number of beams passing through the regions. This will enforce the model to concentrate on the multi-beam overlapped regions to improve the dose prediction accuracy in these regions.

2.4 Model Architecture and workflow

A fully connected 3D U-Net was adopted as the model backbone, taking multi-channels of 3D matrix as input. The 3D input matrix was a concatenation of the pre-processed CT matrices, contour bitmask matrices, and beam mask matrices, depending on the training approach utilized. The model was trained to produce a 3D dose matrix as output based on the ground-truth dose matrix. For clinical testing and deployment, the trained model can be further integrated to Shiva for adaptive proton plan evaluation and potential replanning, as shown in Figure 1.

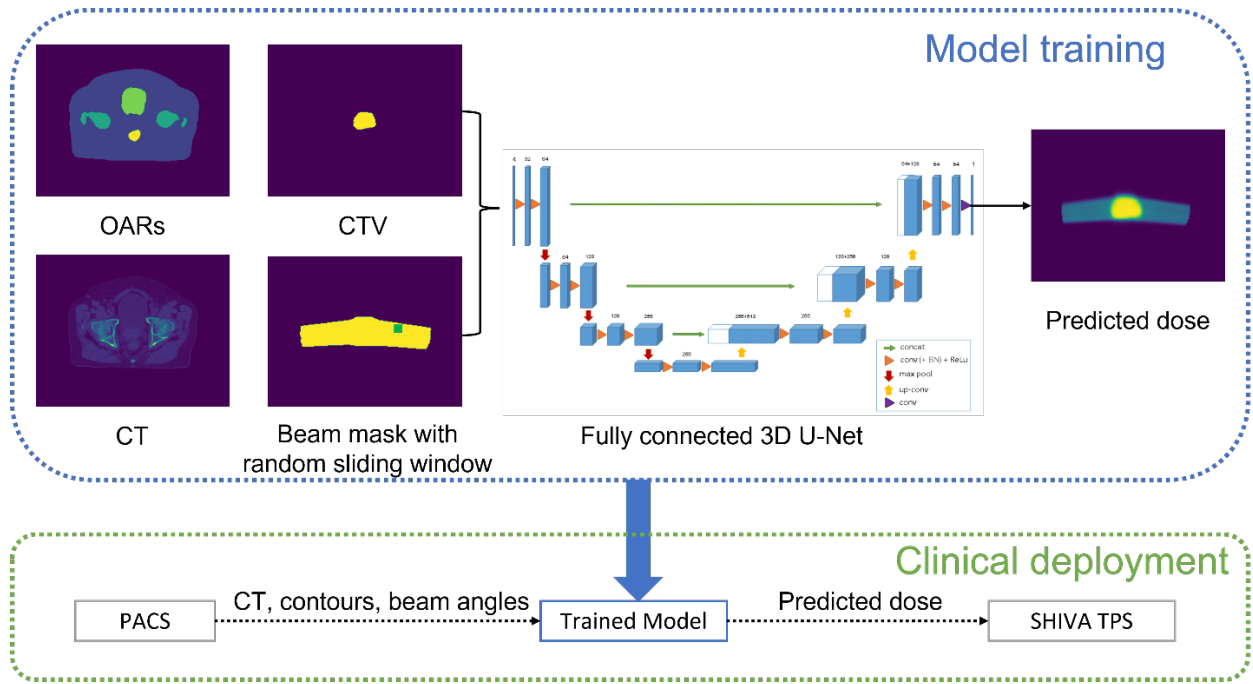

Figure 1. Workflow of the deep learning framework for dose prediction in pencil beam scanning proton therapy.

## 2.5 Enhanced model training with sliding windowing

To improve the model performance, the model training were enhanced using the sliding window technique to concentrate on local details[26]. During the model training, a sliding window with a size of 3×3×3 voxels was randomly selected within the beam mask region and ROI regions. Within the box, the corresponding beam mask values were set to 0, instead of 1. This technique allows the model to concentrate more on local details. The Smooth L1 loss between the predicted dose and the ground truth dose was used as the objective function for the model training as following:

$$L(d, \hat{d}) = \frac{1}{N} \sum_{i=1}^{N} smooth_{L_1}(d_i - \hat{d}_i)$$

in which

$$smooth_{L_1}(x) = f(x) = \begin{cases} 0.5x^2, & if\ |x| < 1 \\ |x| - 0.5, & otherwise \end{cases}$$

where $L$ is the total loss function value, $d_i$ is the ground truth dose at the $i_{th}$ voxel, and $\hat{d}_i$ is the predicted dose at the $i_{th}$ voxel. The Adam optimizer was used to minimize the loss function with the default Reduce-LR-On-Plateau learning rate in PyTorch. To preserve the global information as much as possible, the whole 3D pre-processed matrices as described in the data preprocessing section were used in the input channels for model training as instead of the patch method dividing the whole matrix to many smaller patches during the model training[63]. Random rotation and translation were used to augment the data to avoid overfitting[83]. The models were trained on four NVIDIA Tesla A100 GPUs each with 80 GB on board RAM memory for 200 epochs.

## 2.6 Ablation study

To understand the effectiveness of each component of the proposed strategies, three experiments were designed step-by-step as follows. Experiment 1 used the CT images and contour bitmasks as the input channels for the model training. Experiment 2 added the beam mask into the input channel besides the CT images and contour bitmasks as used in Experiment 1. For Experiment 3, the sliding window strategy was further added to the model training on top of Experiment 2. Ten prostate cancer and 10 lung cancer cases were used to test the performance for each experiment.

In evaluating the accuracy of the dose prediction, we adopted three different evaluation metrics. DVH indices were adopted at first. For DVH index comparison, D2 and D98 for the CTV, Dmean for bladder, spaceOAR, femoral heads, penile bulb, rectum, esophagus, heart, and total lung, and Dmax for spinal cord, spinal cord prv were used. Next, we used the 3D Gamma passing rate with a criterion of 3%/3mm/10% following the AAPM Task Group (TG) report No. 218 recommendation to show the 3D spatial dose distribution agreement between the predicted dose and the ground-truth dose in targets, ROI, and Body (outside ROI), respectively[84]. Finally, we evaluated the 3D spatial dose distribution agreement by the dice coefficients of the structures enclosed by the iso-dose lines (from 10% of the prescription dose to 90% of the prescription dose in an increment of 10%) between the predicted and the ground truth doses. The dice coefficient shows the volumetric similarity between the two volumes.

Table 1. Model training configurations for the conventional ROI model, the beam mask model, and the beam mask with sliding window model

|  | ROI model | Beam mask model | Sliding window model |
| --- | --- | --- | --- |
| Epoch | 200 | 200 | 200 |
| Batch size | 2 | 2 | 2 |
| Training time | 6.9 h | 7.2 h | 7.8 h |
| Predicting time | 0.216 s | 0.228 s | 0.239 s |

# 3. Results

## 3.1 Comparison of dose distributions

Figure 2 shows the distributions of the predicted dose from three different experiments and the ground truth dose in the transversal plane (for prostate) and in the sagittal plane (for lung) of one typical prostate (Fig. 2(a)) and one typical lung (Fig. 2(b)) cancer patient. The dose distribution differences in the corresponding planes between the predicted dose and the ground truth dose of the selected patients were also shown.

## 3.2 Comparison of DVHs

Figure 3 shows the DVHs of the predicted doses from three different experiments and the ground truth doss of one typical prostate (Fig. 3(a)) and one typical lung (Fig. 3(b)) cancer patient. Both figures show that the beam mask model predicted the dose more accurately than the ROI model, while the random sliding window method further improved the dose prediction accuracy than the beam mask method.

Figures 4 shows the absolute deviation of the DVH indices of targets and OARs between the ground truth dose and the predicted doses from three different experiments of the testing cases for prostate and lung. Prostate CTV D98 absolute deviation is 0.53±0.22 Gy[RBE], 0.45±0.25 Gy[RBE], and 0.41±0.27 Gy[RBE] for Experiment 1, 2, and 3 respectively. Lung CTV D98 absolute deviation is 0.74±0.18 Gy[RBE], 0.57±0.21 Gy[RBE], and 0.54±0.15 Gy[RBE] for Experiment 1, 2, and 3, respectively. For prostate and lung CTV D2, the absolute deviation is 0.81±0.25 Gy[RBE] and 0.94±0.44 Gy[RBE] in Experiment 1. The D2 absolute deviation is reduced in Experiment 2 by 0.15 Gy[RBE] and 0.19 Gy[RBE] on average. It is further reduced by 0.1 Gy[RBE] on average in Experiment 3.

For the comparison of the DVH indices of OARs, the predicted DVH index absolute deviation of the Dmean and Dmax between the ground truth dose and the predicted doses from three different experiments of the testing cases for prostate and lung were all within clinical tolerance. For most OARs, the absolute Dmean deviation were around 0.1 Gy[RBE] on average lower in Experiment 2 compared to Experiment 1. Experiment 3 further reduced the Dmean absolute deviation by 0.05 Gy[RBE] on average compared to Experiment 2.

For Dmax of spinal cord and spinal cord prv in lung, the absolute deviation were relatively significant with 0.95±0.19 Gy[RBE] and 1.2±0.27 Gy[RBE] in Experiment 1, respectively. The absolute deviations were reduced to 0.73±0.16 Gy[RBE] and 0.94±0.31 Gy[RBE] in Experiment 2 and further reduced to 0.67±0.15 Gy[RBE] and 0.91±0.28 Gy[RBE] in Experiment 3, respectively.

Moreover, the absolute deviation of Dmean of SpaceOAR, which is an implant used in prostate cancer patients treated with PBSPT to protect the rectum, were 0.51±0.22 Gy[RBE], 0.38±0.17 Gy[RBE], and 0.36±0.12 Gy[RBE] for the three experiments respectively. This shows good dose prediction accuracy even in the implant with further improvements from Experiments 2 and 3.

3.3 3D Gamma evaluation

Figure 5 shows the 3D Gamma passing rates (3%3mm/10%) within targets, OARs, and BODY (outside ROI) of the testing cases for prostate and lung, respectively. Compared to the Gamma passing rates in Experiment 1, the Gamma passing rates improved for target and OARs in

Experiment 2 (prostate targets: 96.93%±0.53% vs. 98.88%±0.49%, lung targets: 93.37%±0.68% vs. 95.31%±0.71%, prostate OARs: 93.21%±0.80% vs. 95.29±0.59%, lung OARs: 90.91%±0.87% vs. 92.7% ± 0.73%). Compared to targets and OARs, the improvements in Gamma passing rates for BODY in Experiment 2 were more remarkable (prostate BODY: 86.88%±0.74% vs. 93.21%±0.56%, lung BODY: 85.14%±0.89% vs. 91.26%±0.61%). The Gamma passing rates were further improved in Experiment 3 by around 1% in targets and by 1.5% in OARs and BODY on average compared to Experiment 2.

3.4 Dice coefficient evaluation

Figure 6 shows the dice coefficients of the structures enclosed by the iso-dose lines (from 10% of the prescription dose to 90% of the prescription dose in a increment of 10%) between the predicted and the ground truth doses of the testing cases for prostate and lung. In general, the average dice coefficients improved for most isodose lines in Experiment 2 compared to Experiment 1. Further improvements were observed in Experiment 3 with a few exceptions: prostate 70% isodose lines between Experiment 1 and Experiment 2, prostate 50% isodose lines between Experiment 2 and Experiment 3, and lung 70% isodose lines between Experiment 2 and Experiment 3, which show comparable dice coefficients. For high percentage isodose lines, slight improvements were observed among three experiments, for example, the 90% isodose lines (prostate 90% isodose lines: 0.959±0.009 vs. 0.971±0.012 vs. 0.979±0.005, lung 90% isodose lines: 0.937±0.029 vs. 0.951±0.011 vs. 0.957±0.023, from the three experiments respectively). For low percentage isodose lines, the improvements among three experiments were remarkable, for example, the 10% isodose lines (prostate 10% isodose lines: 0.895±0.027 vs. 0.927±0.021 vs. 0.945±0.019, lung 10% isodose lines: 0.871±0.027 vs. 0.911±0.023 vs. 0.927±0.017, from the three experiments respectively).

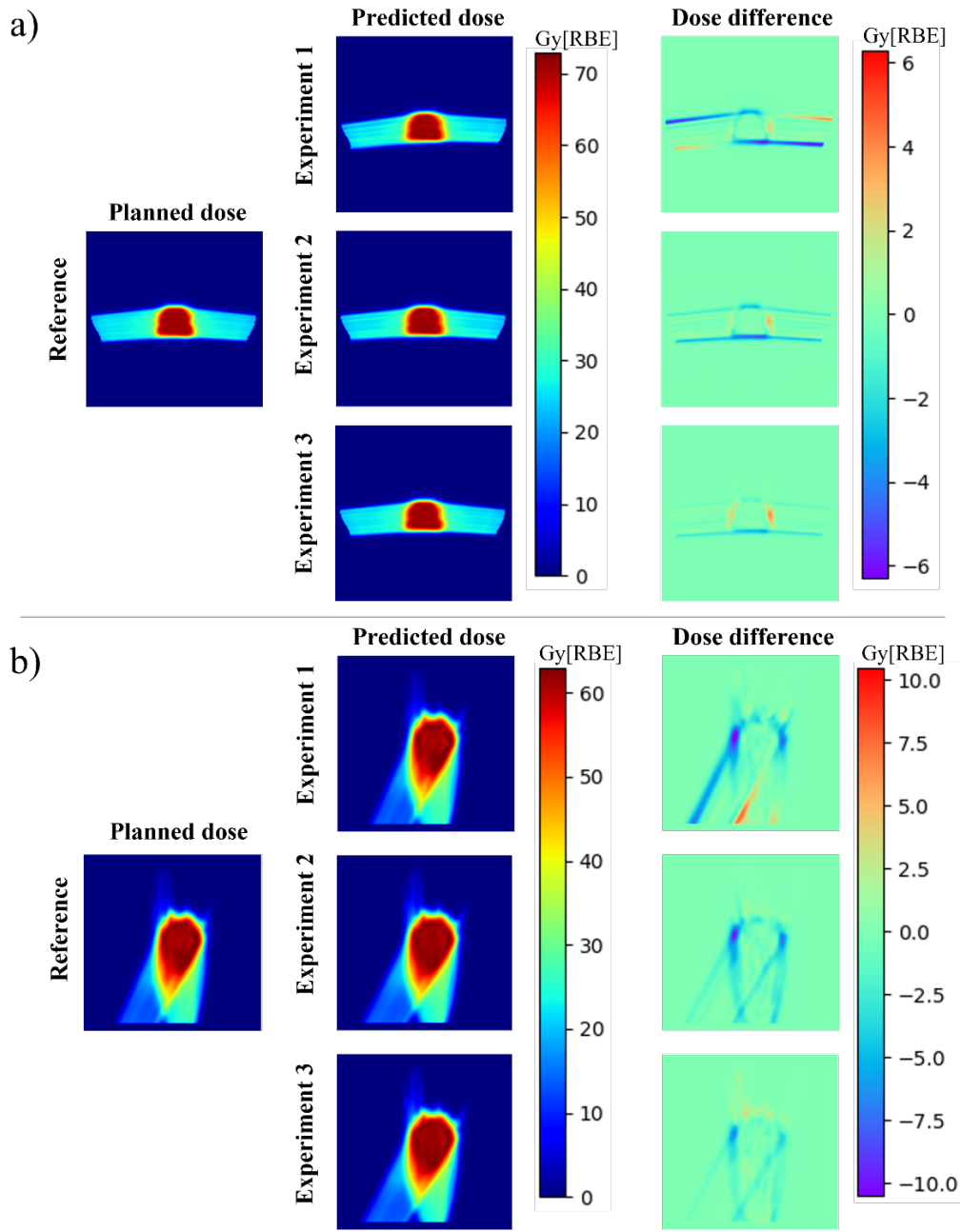

Figure 2. Distributions of the predicted dose from three different experiments and the ground truth dose in the transversal plane (for prostate) and in the sagittal plane (for lung) of one typical prostate (Fig. 2(a)) and one typical lung (Fig. 2(b)) cancer patient

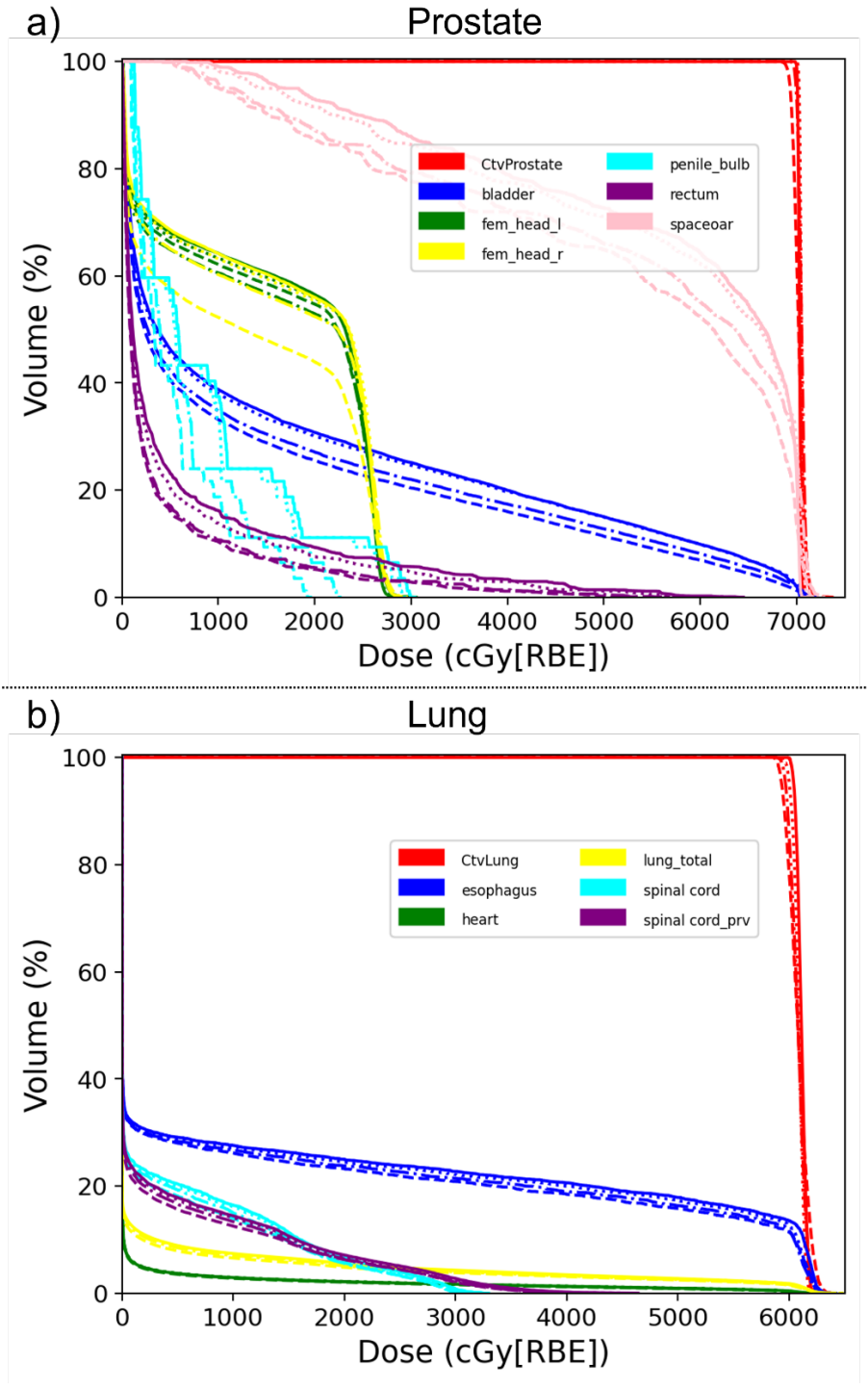

Figure 3. DVHs of the predicted doses from three different experiments and the ground truth dose of one typical prostate (Fig. 3(a)) and one typical lung (Fig. 3(b)) cancer patient. *Solid line*: ground-truth dose; *dash line*: Experiment 1, *dash dot line*: Experiment 2; *dot line*: Experiment 3.

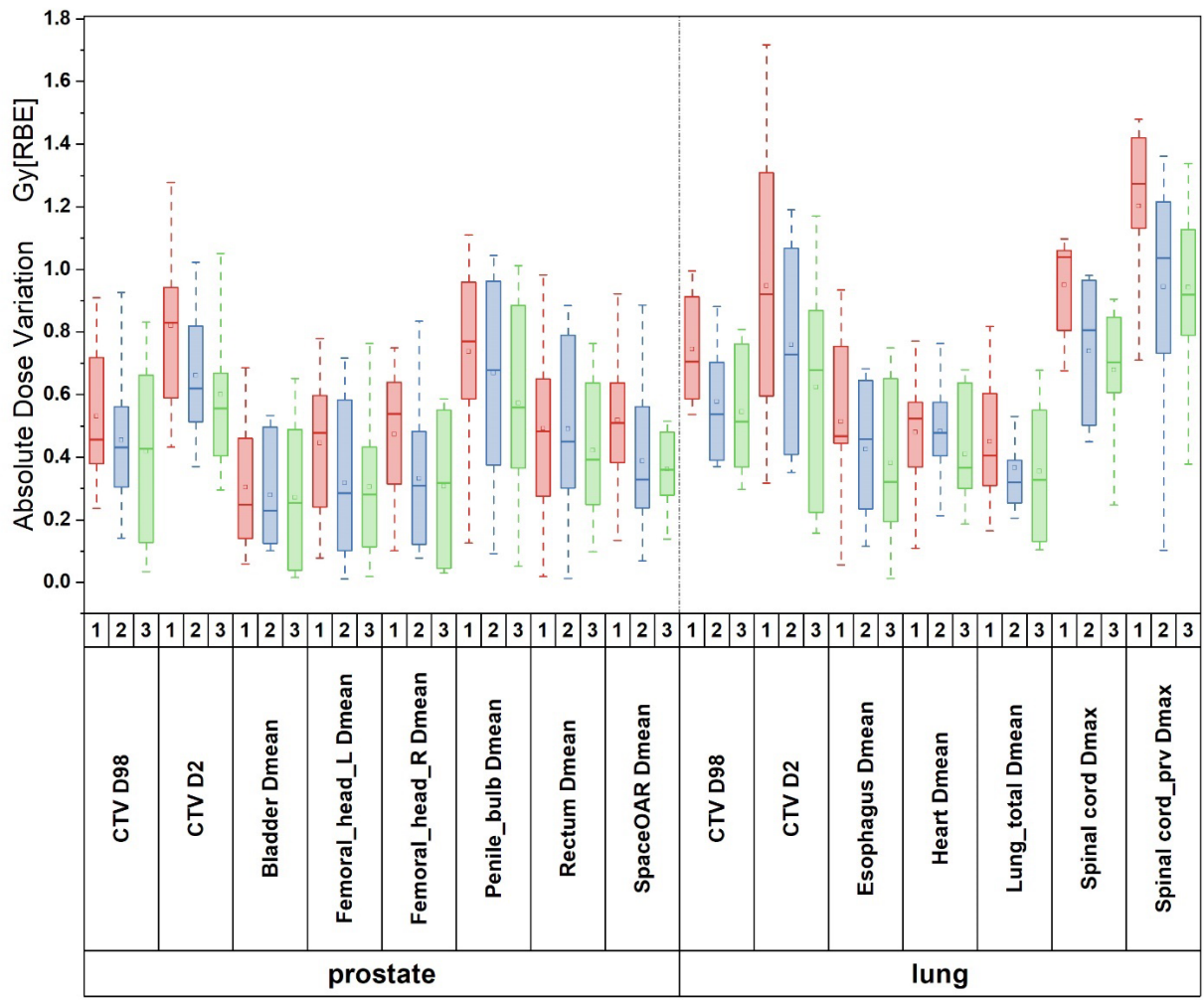

Figure 4. Boxplot (minimum, first quartile, median, third quartile, and maximum, respectively) of absolute deviation of the DVH indices of targets and OARs between the ground truth dose and the predicted doses from three different experiments of the testing cases for prostate and lung. Experiment1, Experiment 2 and Experiment 3 correspond to 1 (*red*), 2 (*blue*) and 3 (*green*) in the figure.

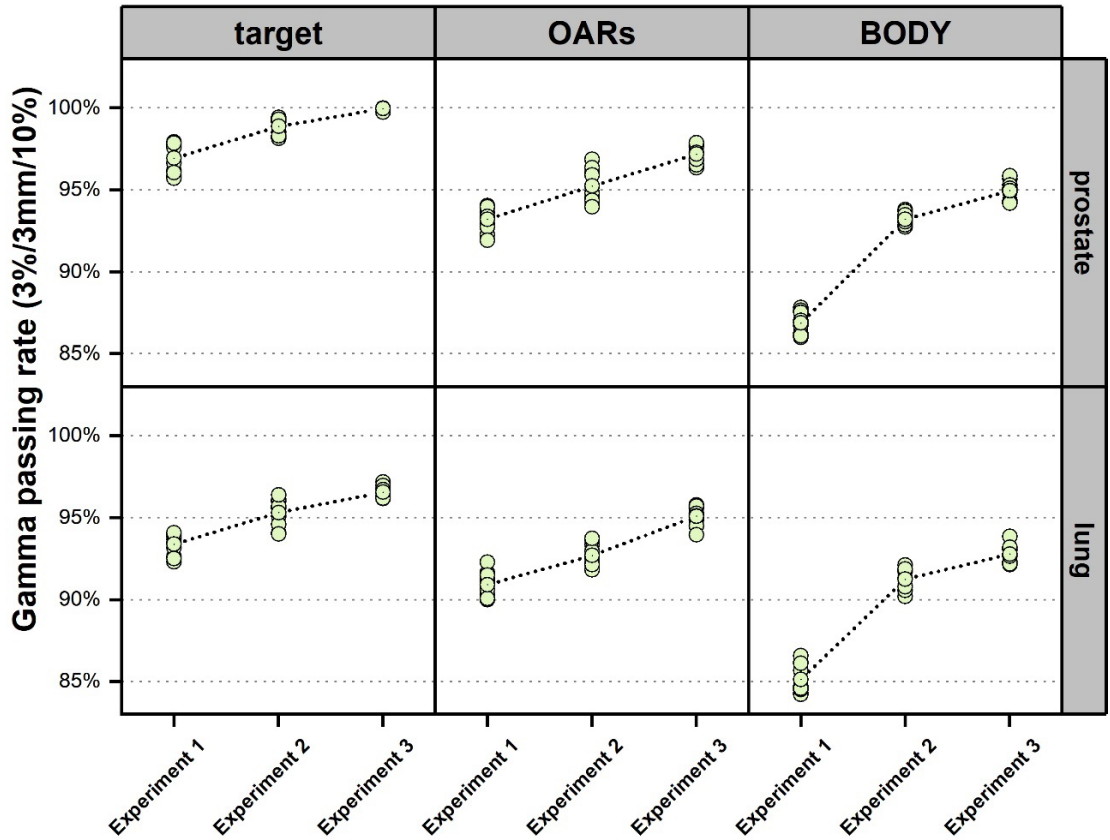

Figure 5. 3D Gamma passing rates (3%3mm/10%) within targets, OARs, and BODY (outside ROI) of the testing cases for prostate and lung, 10 cases from the testing group each (green circles). The dotted lines represent the trend of the average value of all the testing cases by each group.

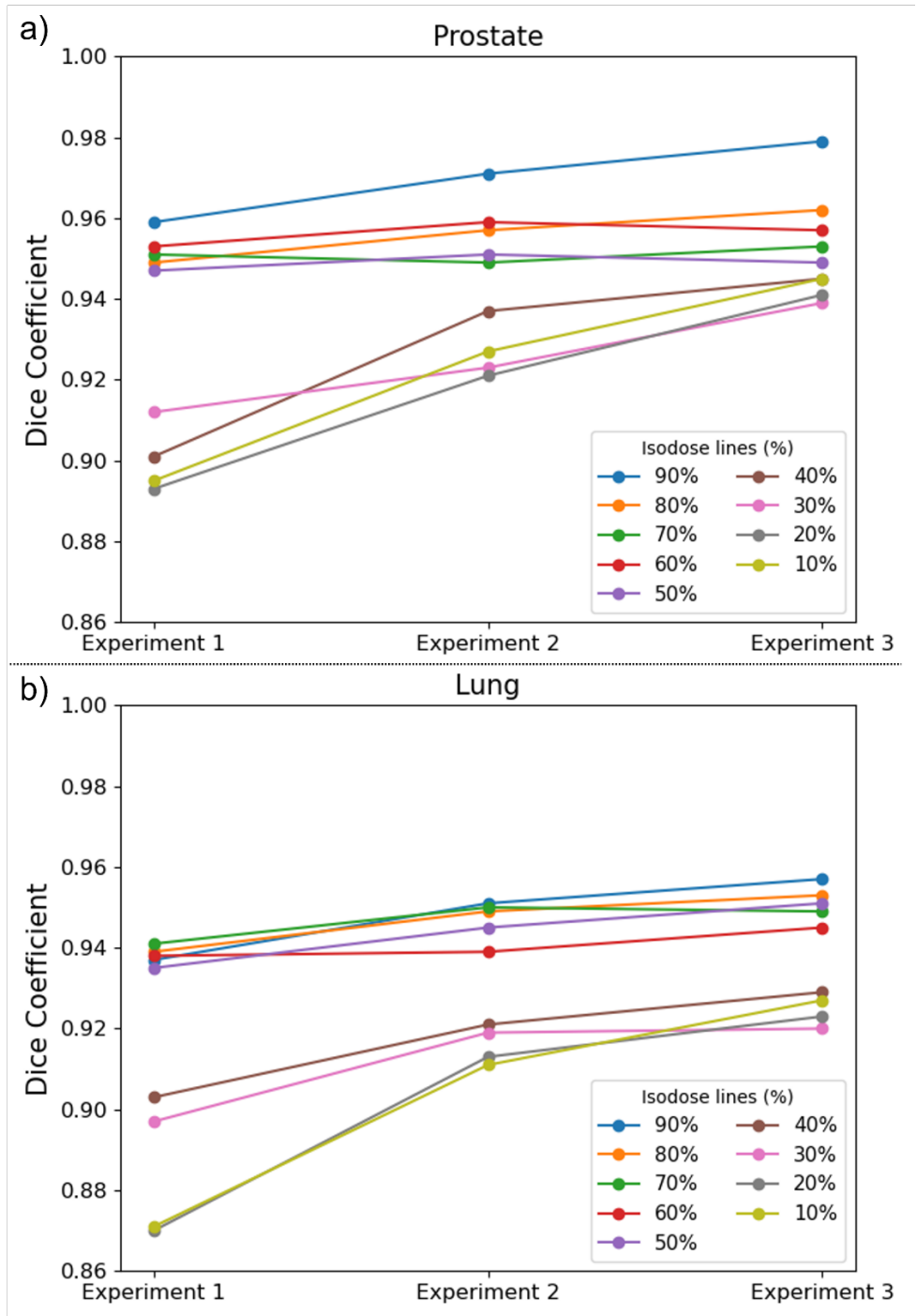

Figure 6. Dice coefficients of the structures enclosed by the iso-dose lines (from 10% of the prescription dose to 90% of the prescription dose in an increment of 10%) between the predicted and the ground truth doses of the testing cases for prostate (Fig. 6(a)) and lung (Fig. 6(b)).

## 4. Discussion

In ART, obtaining a rapid and accurate proton dose distribution based on verification imaging data is critical. This will greatly enhance the efficiency of online ART decision-making, and the subsequent adaptive replanning[75, 85]. Compared to IMRT treatment planning, PBSPT is more sensitive to uncertainties. Additionally, the large number of spots in PBSPT also results in long dose optimization time. To quickly generate sufficiently accurate dose distributions in PBSPT, regardless of the method, remains a challenge. In this study, we introduced the beam mask and the sliding window methods, which greatly improved the conventional ROI method for the dose prediction in PBSPT. Overall, the 3D Gamma passing rates were 99.97±0.07%, 97.18±0.57%, and 95.17%±0.59%. for the testing prostate cases and 96.55±0.34%, 95.09±0.61%, and 92.79±0.53% for the testing lung cases within target, OARs, and BODY, respectively. The dose predictions for all the testing cases were completed within 0.25s. The improved predicted PBSPT doses using our proposed beam mask and sliding window methods met our institutional clinical requirement to have a 3D Gamma passing of 95% with a criterion of 3%/3mm/10%[86-88].

In this study, we introduced the beam mask and sliding window methods to improve the performance of PBSPT dose prediction. For the beam mask method, we observed improvements in DVH index accuracy such as D2, D98, and Dmean, for both targets and OARs with bigger improvements in spinal cord Dmax. The introduction of the random sliding window further improved the DVH index accuracy, demonstrating the effectiveness of these two methods from a clinical perspective. It is worth noting that for some prostate patients with implants, such as the SpaceOAR, our proposed methods could accurately predict Dmean of these structures. This suggests that, in addition to predicting the dose distributions of human tissue structures, the model also performs well in predicting the dose to non-human tissue implants.

To further illustrate the 3D spatial dose distribution agreement between the predicted dose and the ground-truth dose, we used 3D Gamma evaluation with a criterion of 3%/3mm/10%. In the past literatures based on the ROI, usually only the dose related metrics within ROI were reported[61-66, 73-75]. However, the dose prediction accuracy outside ROI is also clinically important, especially for the identification of possible hot spots outside the ROI and the protection of important OARs not included in the ROI due to various reasons such as omission, mistakes, or limitations due to the TPS. This is a significant limitation of prior studies since, although the dose prediction accuracy within targets is clinically relevant, it is more trivial for the DL models to predict the dose distribution within targets since the optimized dose distribution is approximately uniform within targets. In terms of evaluating the DL models, the dose prediction accuracy outside the ROI is more indicative of the model's performance. One possible reason for poor performance outside the ROI in prior studies might be that by using the ROI as input, the DL model emphasizes more on the ROI but less on the regions along the beam path outside the ROI.

To avoid this possible bias, we evaluated the 3D Gamma passing rates in three regions: targets, OARs, and BODY (outside ROI). The beam mask and sliding window methods significantly improved the 3D Gamma passing rates for both the testing prostate and lung cancer patients in all regions compared to the conventional ROI method. Relative to the targets and OARs, our methods greatly improved the Gamma passing rates in the BODY, outside of ROI. The use of the beam mask allows the model to be exposed to more beam path information, resulting in a more accurate dose distribution along the beam path. Furthermore, the random sliding window method led to higher loss in small local areas during the model training, thereby steering the model to pay more attention to local details. The combination of these two methods led to a significant increase in the Gamma passing rates in targets, OARs, and particularly BODY (outside the ROI). The dice coefficients, reflecting the volumetric similarity of the structures enclosed by the iso-dose lines (from 10% of the prescription dose to 90% of the prescription dose in an increment of 10%) between the predicted and the ground truth doses, further prove this point. Our proposed methods have high dice coefficients in the high percentage isodose lines-enclosed regions, demonstrating good performance of the proposed methods within targets and the target penumbra regions. In the low percentage isodose lines-enclosed regions, such as 10%, our methods significantly improved the corresponding dice coefficients, indicating substantial improvement in the dose prediction accuracy outside the ROI. These results suggested that our proposed methods can achieve clinically accurate results not only in the ROI, but also outside the ROI. This is of great clinical significance for online adaptive proton therapy in clinical decision making and subsequent replanning.

Recently, several studies have been published on proton dose prediction. For example, some studies focused on DL-based proton dose calculation engine development considering basic physics to accelerate the proton dose calculation[68-70]. Another direction was to convert dose distributions calculated by the pencil beam algorithm in proton therapy to more accurate dose distribution calculated by a Monte Carlo dose engine with deep-learning methods[71]. Some studies have attempted to use the ROI method to train DL models to predict PBSPT plan dose distributions[73]. While these methods had achieved acceptable accuracies in the targets, they often showed considerable deviations in certain OARs, especially outside the ROI. Subsequent studies used the modulated spots weights optimized using a pencil beam-based dose engine as the input channel and converted the dose distribution derived by a pencil-beam-based dose engine to the more accurate dose distribution derived from MC-based methods[74]. Though the accuracy was improved, this approach required extensive computation and time for pencil-beam-based optimization to get the spots weights. This limited its clinical applications, especially in time-sensitive online adaptive proton therapy. Another recently published study adopted a different approach, also using the ROI method for dose prediction in PBSPT. While the Gamma passing rate (3%/3mm/10%) for the BODY was only around 85%, a manual post-processing method was employed to replace the specific voxel dose in the predicted dose distribution, making the corresponding dose distribution within the ROI more clinically reasonable. The post-processed dose was then used for the dose mimicking based plan generation[75]. Although the study demonstrated that the final deliverable plans were clinically acceptable in the test cases, the postprocessing method used in the dose prediction process raised concerns about physics interpretability and accuracy in clinical scenarios where specific voxel doses were manually replaced. From a clinical perspective, it is more accurate, efficient, and acceptable to obtain a dose distribution close to the ground truth dose distribution directly rather than post-processing a not-so-accurate dose distribution.

Our proposed method balanced model accuracy and complexity. The use of beam masks and sliding windows remarkably improves the dose prediction both inside and outside the ROI compared to the ROI method. Moreover, the generation of the beam masks only requires beam angles, without the need for complex beamline and spots weight calculations. The random sliding window method is straightforward to implement, making it simpler to deploy the developed software in clinical scenarios, particularly in adaptive radiotherapy.

Since all the PBSPT plans used for the model training and testing were calculated by the Monte Carlo (MC)-based dose engine, we can achieve dose distribution with an accuracy comparable to the ones calculated by the Monte Carlo-based dose engine in prostate and lung cancer patients treated with PBSPT. Moreover, the dose predictions for all the testing cases were completed within 0.25s, which is much faster than any existing MC-based dose engines, even those that utilize GPU acceleration for MC calculations. As a result, our methods make real-time plan dose prediction of PBSPT plans possible with comparable accuracy as MC-based methods.

It is important to note the limitations of this study. While the model provides highly accurate dose predictions for prostate cancer patients, its accuracy for lung cancer patients, although clinically acceptable, is relatively lower in comparison. This is due to the tissue heterogeneities in thoracic disease site, which is often more challenging for dose calculation in PBSPT. Additionally, the beam configurations for lung patients are more varied than those for prostate patients. In our subsequent studies, we will collect more lung cancer patients to further fine tune the DL model, improving its accuracy in lung cancer. However, considering that prostate cancer patients account for nearly 40% of patients treated at our proton centers, this model holds immense value in reducing the workload at our center and the proton centers, which treat lots of prostate cancer patients.

## 5. Conclusion

In summary, we have proposed an improved PBSPT dose prediction method utilizing the beam mask and the random sliding window methods. This method predicts dose distributions that closely resemble the ground truth diose distributions for both prostate and lung cancer patients, both within and outside the ROI. It strikes an optimal balance between the model accuracy and complexity, providing accurate dose predictions without the need for complicated pre-calculated energy layer and spots weight information, making it easy to be implemented in routine clinical use. This dose prediction method can be utilized for online adaptive proton therapy in clinical decision making and subsequent replanning.